# Projections for COVID-19 spread in India and its worst affected five states using the Modified SEIRD and LSTM models

Punam Bedi, Shivani, Pushkar Gole, Neha Gupta, Vinita Jindal

## Abstract

The last leg of the year 2019 gave rise to a virus named COVID-19 (Corona Virus Disease 2019). Since the beginning of this infection in India, the government implemented several policies and restrictions to curtail its spread among the population. As the time passed, these restrictions were relaxed and people were advised to follow precautionary measures by themselves. These timely decisions taken by the Indian government helped in decelerating the spread of COVID-19 to a large extent. Despite these decisions, the pandemic continues to spread and hence, there is an urgent need to plan and control the spread of this disease. This is possible by finding the future predictions about the spread. Scientists across the globe are working towards estimating the future growth of COVID-19. This paper proposes a Modified SEIRD (Susceptible-Exposed-Infected-Recovered-Deceased) model for projecting COVID-19 infections in India and its five states having the highest number of total cases. In this model, exposed compartment contains individuals which may be asymptomatic but infectious. Deep Learning based Long Short-Term Memory (LSTM) model has also been used in this paper to perform short-term projections. The projections obtained from the proposed Modified SEIRD model have also been compared with the projections made by LSTM for next 30 days. The epidemiological data up to 15[th] August 2020 has been used for carrying out predictions in this paper. These predictions will help in arranging adequate medical infrastructure and providing proper preventive measures to handle the current pandemic. The effect of different lockdowns imposed by the Indian government has also been used in modelling and analysis in the proposed Modified SEIRD model. Moreover, this paper predicts the impact on the pandemic spread when these restrictions are enforced further. These projections will act as a beacon for future policy-making to control the COVID-19 spread in India.

## 1. Introduction

The world witnessed the COVID-19 (Corona Virus Disease 2019) pandemic that affected the lives of many people across the globe. The spread of this virus has taken an exponential speed. Therefore, on 11[th] March 2020, the World Health Organization (WHO) announced COVID-19 as a "global pandemic" [1]. Due to the outbreak of COVID-19, an unavoidable situation was created, whereby the authorities of various countries and continents had to put restrictions on the movement of people and non-essential activities. Some of these restrictions included imposing of lockdowns, maintaining social distancing, work from home in academics and in business continuity plans. Thus, the spread of COVID-19 has left a major impact on the environment as well as on the lifestyle of human beings [2], [3]. Almost all educational institutions were closed, sports leagues were cancelled, and people were advised to work from home, and perform contactless financial transactions using various digital platforms [4], [5].

COVID-19 was first reported in the Wuhan city of China on 17th November 2019 and from there it has spread to the whole world [6], [7]. Due to the spread of this disease through human to human contact, a large number of cases have been reported worldwide [8]. Recent studies have found that a healthy person can be infected by coming in contact with an infected person or with the surface touched by an infected person to which the Coronavirus got transferred. Also, the symptoms of Coronavirus in an infected person are visible after a certain time period. During this time period, the infected person is a carrier of Coronavirus and is able to infect other healthy persons. As of 16th August 2020, more than 21 million people have been infected and more than 0.7 million people have died from COVID-19 across the globe [9]. Therefore, COVID-19 has become a big threat to people and environment [10], [11].

To tackle this difficult situation, the first step is to take precautionary measures to prevent the infection and the second step is that infected people must quarantine themselves and get medical help. Taking precautions on an individual level is also required, such as use of sanitizer, use of face mask and maintaining social distancing. These resources namely sanitizers and face masks are the need of time [12]. However, it is equally important to properly dispose-off the used masks so as to protect the environment. Governments have also taken many steps to arrange necessary resources to provide better medical services to the infected people. An estimate of these resources is created so that all the needs of people can be met in time. Different methods are being used by researchers to estimate the resources.

For a country like India, having a large population of around 1.38 billion, it is a challenging task to handle this pandemic efficiently [13]. In India, the first COVID-19 positive case was reported in the Thrissur district of Kerala on 30th January 2020, of a student who returned back from Wuhan university, China. During the initial period, there wasn't a substantial increase in the number of cases in India and by 15th March, the number of cases barely crossed the figure of 100. As of 16th August 2020, COVID-19 has spread to 215 countries with more than 6 million active infected cases globally. In India, the number of COVID-19 cases has crossed 2.5 million with more than 0.6 million active infected cases, 1.8 million recovered cases and 50,122 as the total number of deaths, which makes India the worst affected country in Asia. The variation in the total number of infected cases is evident in India with highest reported cases from Maharashtra and lowest from Mizoram [14].

The number of cases globally as well as in India is increasing at a very rapid rate. As is evident from the data, state Maharashtra is the worst affected in terms of total cases which accounts for about 23% of the cases in India. The next four worst affected states/ union territories are Tamil Nadu, Andhra Pradesh, Karnataka and Delhi having approximately 38% of the total cases and the rest of Indian states/ union territories having another 39% cases. North eastern states of India are much better like Mizoram, Sikkim and Meghalaya, each having less than 1500 cases so far. To understand the future spread of pandemic and to devise management strategies, various models have been designed, which give information regarding the time of attainment of infection peak, the number of infected cases and the requirement of medical infrastructure to manage the spread [15], [16].

In this paper, an epidemiological model named Modified SEIRD (Susceptible-Exposed-Infected-Recovered-Deceased) model has been proposed. It utilises the real data of infections, recoveries and deaths caused by COVID-19 to make predictions. We have used the data of India and its five states having the highest number of total cases to make predictions using Modified SEIRD model. Considering the data up to 15$^{th}$ August 2020, Maharashtra, Tamil Nadu, Andhra Pradesh, Karnataka, and Delhi are the five states with the highest number of reported cases. The proposed model uses a parameter named epsilon for COVID-19 projections, which takes into account the proportion of Exposed population that is asymptomatic but infectious, and is unknowingly spreading the infection. This paper predicts the number of cases in Infected, Exposed, Recovered and Deceased compartments. Student t-test was used to obtain confidence levels for time-series data in consideration [17]. T-test is useful when the sample size is very small as compared to large population size. Since the data available is limited, t-test becomes an appropriate choice to find the confidence intervals for COVID-19 predictions. The effect of different lockdowns imposed by the Indian government has also been utilized in modelling using the proposed Modified SEIRD model.

Furthermore, this paper utilises Deep Learning (DL) based Long Short-Term Memory (LSTM) model to perform short-term projections for the next 30 days i.e. from 16$^{th}$ August 2020 to 14$^{th}$ September 2020. The results obtained by LSTM model have been compared with the results of the Modified SEIRD model. The epidemiological data up to 15$^{th}$ August 2020 has been used to obtain projections in this paper. The upper and lower estimates of the predictions made by both the models have also been calculated using 90% confidence intervals. The same has been shown in the graphs corresponding to short-term predictions. It has not been shown for long-term predictions because the values are very close to the reported data and are not distinguishable on the scale chosen for these graphs.

The rest of the paper is organized as follows: the next section presents the Review of Literature. In Section 3, various models used for projections, namely, the proposed Modified SEIRD model along with SEIR, SEIRD models and the LSTM have been explained. Section 4 presents the Experimental Setup. The Results are discussed in Section 5, which is followed by Conclusion at the end.

## 2. Review of Literature

COVID-19 is a communicable disease that has been declared as a pandemic by the WHO [18]. Moreover, there is no medicine or vaccine available to cure this infection as of now. Hence, the only way to protect oneself from this pandemic is to get protected from the contact of any infected person. With the ongoing pandemic threat, researchers started to study the future of COVID-19. These research groups are mainly divided into two categories, where one tries to find the vaccine and other group tries to predict the damage that can be caused by this disease. Based on the predictions, resources can be prepared to treat people and minimize the fatalities. This paper predicts the future trend of COVID-19 by modelling the effect of different lockdowns imposed by the Indian government.

Since the beginning of COVID-19, various researchers have predicted its spreading trend for different countries and their states [19], [20]. Ahmed [21] has studied the effect of patient age, gender and their geographical location on the infection spread. In his work, the population of India that has arrived from different regions is taken into consideration. These people were divided into six groups to study the regional effect of Corona on a patient. Clustering and Multiple Linear Regression are two techniques that have been used for this study. Clustering has been used to find the similarity between different groups. Multiple Linear Regression has been used to predict the source of infection by assigning the new patient case into one of the above defined groups. In every group, different age distributions have been studied and their recovery rates were calculated.

Ceylan [22] utilised Auto-Regressive Integrated Moving Average (ARIMA) models to predict the future trend of the Coronavirus disease in the three worst affected countries of Europe, namely Italy, Spain and France. The author formulated several ARIMA models using different parameter values. The best models were then used for estimating the spread of the disease in each of the three countries. Patrikar et al [23] have used the modified SEIR method in their work to predict the curve of COVID-19 for India. In this modified SEIR, the effect of social distancing has been studied on the COVID-19 and different graphs have been obtained. It was also concluded that social distancing is working in India, according to the model given by Poonia et al [24]. In the paper, authors performed short term forecast of COVID-19 for different states of India in worst case scenario.

SEIR model was also used to study the effect of temperature on COVID-19 outbreak in China [25]. In the paper, the authors incorporated climatic factors in the original SEIR model to analyse their impact on the spread of COVID-19. Due to the dynamic nature of the Coronavirus spread, many researchers have performed short-term forecasting of the future spread of this deadly virus. Roosa et al [26] performed 5-day and 10-day forecasts of the total confirmed cases for Guangdong and Zhejiang provinces of China. The authors used logistic growth model, Richards growth model and a sub-epidemic wave model to generate predictions. A bootstrap approach was adopted in the paper to compute the uncertainty bounds for predicting cumulative cases in near future.

Arora et al [27] made use of different variations of Long Short-Term Memory models for short-term prediction of COVID-19 cases in India. Deep LSTM, Bi-directional LSTM (Bi-LSTM) and Convolutional LSTM (Conv-LSTM) were used to calculate one-week predictions for different states and union territories of India. Tomar et al [28] also used LSTM along with power-law curve fitting to predict the trend of COVID-19 in India. The authors forecasted the total number of confirmed, recovered and deceased cases for a short time span of next 30 days. In addition, the paper also analysed the effect of different values of transmission rate on the number of predicted infected cases.

Pal et al [29] combined the COVID-19 data with the weather statistics of different countries to predict active cases using shallow LSTM model. The authors used Bayesian optimization together with fuzzy rules to predict the future risk of Coronavirus in their work. Pandey et al [30] analysed the initial outbreak of COVID-19 in India and forecasted the trend for two weeks in future. The authors used the SEIR model to predict the future trend of the virus in India, with and without interventions. Additionally, Regression model was also used by the authors to predict the change in the number of confirmed and death cases in India.

Chakraborty et al [16] proposed a hybrid approach to generate ten-day predictions for United Kingdom, France, India, Canada and South Korea. The model combined Wavelet model and ARIMA models for computing predictions for next ten days. Various studies have been recently conducted by the researchers to understand the dynamics of this pandemic for different regions across the globe [3], [1], [31], [32], [33], [34]. The forecast of COVID-19 in Indian context also has been investigated by many researchers using mathematical and epidemiological models, but limited contribution exists for its states [35], [19], [36], [37], [38], [23].

In this paper, a compartmental epidemiological model, named Modified SEIRD (Susceptible-Exposed-Infected-Recovered-Deceased) model has been proposed for the projection of COVID-19 spread in India and its five states with the highest number of total cases. Further, short-term predictions for next 30 days have also been computed using the LSTM model and the results are compared with the projections of the proposed Modified SEIRD model. The forecasts are as good as the quality of data available and, therefore, the future spread of the virus may also affect the predictions. The next section presents the proposed Modified SEIRD model and the LSTM model used for projections.

## 3. Models used for Projections

This section describes the proposed Modified SEIRD model based on SEIR and SEIRD epidemiology models, and the Long Short-Term Memory Model (LSTM) model, that have been used in this paper for projecting the trend of COVID-19 in India and its five states having the highest number of total cases.

### 3.1. Proposed Modified SEIRD model

The proposed SEIRD model for COVID-19 is based on SEIR and SEIRD epidemiology models which are first described below.

**SEIR and SEIRD Models**

To model the trend of a disease, researchers have used various epidemiology models in the past [35], [38], [39]. One of the most common models used in the literature is SEIR (Susceptible-Exposed-Infectious-Recovered) model [34] [40]. It is a mathematical data modelling technique based on SIR (Susceptible-Infectious-Recovered) model, which is used for forecasting the spread of an epidemic [41], [42]. The SEIR model is diagrammatically represented in Figure 1.

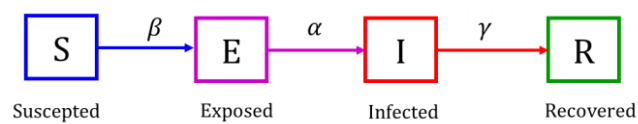

Figure 1: SEIR Model

In SEIR model, the population is divided into four compartments namely S, E, I, and R [43], [44]. Each of these compartments can take different values with respect to time, representing their dynamic behaviour. These have been described as follows:

- S(t) - The number of individuals who are susceptible to the disease, i.e. who are not (yet) infected on day t.
- E(t) - The number of individuals who have been in contact with the infected people and are exposed to the disease on day t, but disease symptoms are not yet visible in them. Such individuals are called asymptomatic.
- I(t) - The number of infected individuals on day t, assumed to be infectious and are able to transmit the infection to others.
- R(t) - The number of individuals who were infected, but they have recovered on day t and developed immunity. This also includes the people who have died [45], [33], [46]. Some authors name this compartment as Removed compartment.

The assumptions related to the SEIR model are listed below:
- There is no entry or departure from the population except possibly through death from the disease [35].
- Recovered people become immune to the disease, and can no longer spread the infection.
- The increase in the number of infected people is directly proportional to the number of infected people as well as to the size of the exposed population.
- Number of recovered persons is directly affected by the number of persons being infected.

In SEIR model, shown in Figure 1, the transmission rate β, regulates the flow of spread which describes the possibility of spreading infection within a susceptible and infectious individual. It represents the average contact frequency. $\alpha$ represents the onset rate where $1/\alpha$ is average latent period. Infected individuals leave the infected compartment at a rate $\gamma I$ to join the recovered class, $1/\gamma$ being the average infectious period.

The spread of a disease may lead to many casualties. However, this category of people is not included in SEIR model separately. Therefore, to represent this group of people separately, a new compartment 'D' is added to the existing SEIR model and the resulting model is known as SEIRD model [47], [48]. SEIRD model has been depicted in Figure 2. Here, D(t) denotes the number of deceased individuals on day t. Infected people become Deceased with a rate $\mu I$.

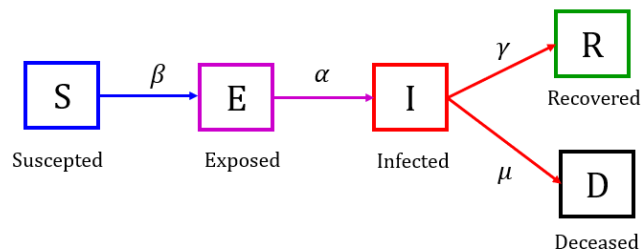

Figure 2: SEIRD Model

As the COVID-19 is a pandemic and SEIRD model works well for such situations, so this paper uses the SEIRD model and proposes SEIRD model with infectivity in exposed population to make predictions. The proposed Modified SEIRD model has been described in the next sub-section.

## Proposed Modified SEIRD model

SEIRD model assumes that exposed population is non-infectious, whereas it has been seen in COVID-19 cases that asymptomatic individuals are tested positive and are also responsible for spreading the disease. Hence, we have modified the SEIRD model to include infectivity in exposed population. The SEIRD model has been modified with the introduction of parameter 'ε' (epsilon), which accounts for the part of Exposed population that is asymptomatic, but infectious, and hence infecting others [35]. The proposed model is named as Modified SEIRD model and is shown in Figure 3.

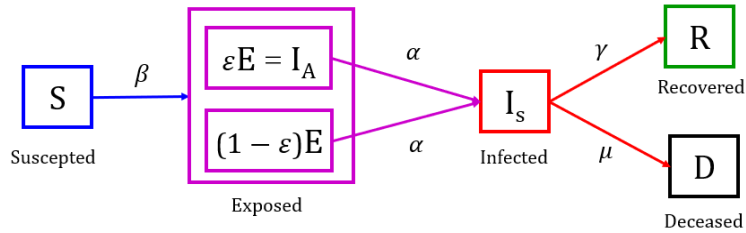

Figure 3: Modified SEIRD Model

The Modified SEIRD model is governed by the following differential equations:

- The susceptible equation: $\frac{dS}{dt} = -\frac{\beta S(I+\varepsilon E)}{N}$ (1)
- The exposed equation: $\frac{dE}{dt} = \frac{\beta S(I+\varepsilon E)}{N} - \alpha E$ (2)
- The infected equation: $\frac{dI}{dt} = \alpha E - \gamma I - \mu I$ (3)
- The recovered equation: $\frac{dR}{dt} = \gamma I$ (4)
- The deceased equation: $\frac{dD}{dt} = \mu I$ (5)

Here, N = S + E + I + R + D, is the total population. Each person belongs to one of the five compartments namely Suscepted, Exposed, Infected, Recovered and Deceased. A person can shift from one compartment to other. The parameters used in equations (1), (2), (3), (4) and (5) regulate the shift of people among different compartments. These parameters are discussed below:

- Beta (β): This parameter represents the transmission rate per capita. It denotes the number of persons which come in contact with an infected person per day.
- Epsilon (ε): This parameter denotes the proportion of exposed people which are infectious and unknowingly infecting other susceptible people.
- Alpha (α): It represents the onset rate where $1/\alpha$ is average latent period.
- Gamma (γ): It represents the recovery rate where $1/\gamma$ is mean infectious period.
- Mu (μ): This parameter denotes the rate at which infected people become deceased.

To study the growth of infection, researchers have used the basic reproduction number, R0, and effective reproduction number, Re, as evaluation metrics in their work [49]. Basic reproduction number Ɍ0 is defined as the average number of secondary infections generated when one infected person is introduced into a host population where everybody is susceptible. Ɍ0 is calculated by equation (6):

$$Ɍ0 = \left(\frac{β}{(γ+μ)}\right) + ε * \left(\frac{β}{α}\right) \tag{6}$$

Effective Reproduction number, Ɍe, is the average number of new infections generated by an infectious individual on day t, in the partially susceptible population S. It is calculated by equation (7):

$$Ɍe = Ɍ0 * \frac{S}{N} \tag{7}$$

Equations (6), (7) clearly show the increase by a factor $ε * \left(\frac{β}{α}\right)$ in the value of Ɍ0 and the corresponding increase in Ɍe by the exposed infectious population. Hence the parameter ε, representing the proportion of infectious exposed population contributes to the growth of COVID-19. High value of this parameter indicates that there are more infectious exposed persons and contact tracing should be done, so that these persons can be isolated to reduce the spread of the disease.

## 3.2. Long Short-Term Memory (LSTM) Model

Deep Learning (DL) is a branch of Machine Learning which is inspired by the working of the human brain. Some of the popular DL techniques are Deep Neural Network (DNN), Convolutional Neural Network (CNN), Recurrent Neural Network (RNN) etc. RNNs are an extension of DNNs that consist of feedback links along with feed-forward connections [50]. Unlike a DNN, RNNs use the previous output to compute the next output. So, they can efficiently process natural language, recognize speech, and perform image captioning as compared to other DL techniques [51].

RNN suffers from the problem of vanishing and exploding gradient [52]. To solve these problems, LSTM networks are used. LSTM is an extension of RNN which allows the network to learn long-term dependencies in the input data. It processes and forecasts time-series data very efficiently. Each LSTM cell comprises of three gates: Input Gate, Forget Gate, and Output Gate. Figure 4 depicts these three gates present in each LSTM cell.

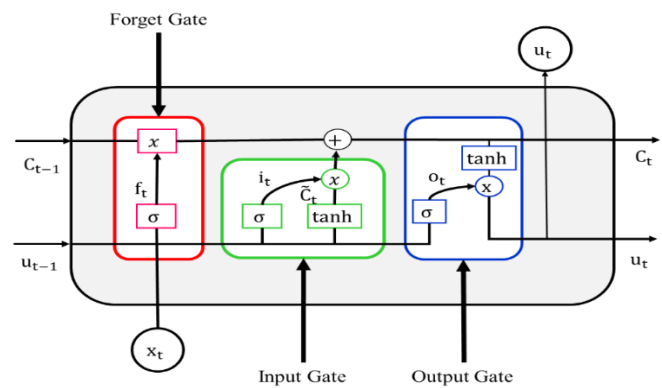

Figure 4: LSTM cell

The equations (8), (9), (10), (11), (12) and (13) are used in LSTM cell. $W_i, W_C, W_f, W_o$ and $b_i, b_C, b_f, b_o$ are the weights and biases, respectively.

$$i_t = \sigma(W_i \cdot [u_{t-1}, x_t] + b_i) \tag{8}$$

$$\tilde{C}_t = \tanh(W_C \cdot [u_{t-1}, x_t] + b_C) \tag{9}$$

$$f_t = \sigma(W_f \cdot [u_{t-1}, x_t] + b_f) \tag{10}$$

$$o_t = \sigma(W_o \cdot [u_{t-1}, x_t] + b_o) \tag{11}$$

$$C_t = f_t C_{t-1} + i_t \tilde{C}_t \tag{12}$$

$$u_t = \tanh C_t \times o_t \tag{13}$$

In each LSTM cell, tanh activation function is used, which distributes the gradients computed in the backpropagation algorithm while training the LSTM network [51]. Hence, it solves the vanishing and exploding gradient problem of RNN.

As the data of COVID-19 is a time-series data and LSTM models time-series data very well, so we have used LSTM network to model the COVID-19 data of India and its five states having the highest number of total cases.

## 4. Experimental Setup

In this paper, the experimentation was performed in Python using Jupyter Notebook and PyCharm Integrated Development Environment. COVID-19 data was collected for India as well as its five states having highest number of total cases. The epidemiological data till 15[th] August 2020, was taken from the *covid19india.org* website [14]. For India, the data for Confirmed, Recovered and Deceased cases is available since 30[th] January 2020. For states, the data is available 14[th] March 2020 onwards. In case of India, data is available for Daily Confirmed, Total Confirmed, Daily Recovered, Total Recovered, Daily Deceased and Total Deceased cases. So, no pre-processing was required for the data of India. However, in case of states, only Daily Confirmed, Daily Deceased and Daily Recovered cases were available. So, pre-processing was performed on this data to obtain Total Confirmed, Total Deceased and Total Recovered cases for all states. These have been calculated by taking the cumulative sum of Daily Confirmed cases, Daily Deceased cases and Daily Recovered cases respectively.

For experimentation, two different models based on epidemiological and DL have been used to make projections. In this paper, the proposed Modified SEIRD, an epidemiological model and LSTM, a deep learning-based model, have been used. Further, t-test has been used to compute the lower and upper estimates of 90% confidence interval for the predictions and are shown in Figure 11 (a-f) -16 (a-f) for short-term predictions. The experimental setup for both of these models have been described in the upcoming sub-sections.

## 4.1. Experimental Setup for the proposed Modified SEIRD model

For experimentation, proposed Modified SEIRD model uses various parameters mentioned in Section 3 above. The values of these parameters, β, ε, α, γ and μ, have been estimated in this paper through curve fitting to actual data, performed using different functions of Python's *lmfit* library. The optimized values of these parameters have been obtained by the *minimize* function. This function uses dual annealing method for obtaining a global optimal solution. The initial value of infected, I0, is taken as 1 based on the assumption that the infection started from one person. The initial value of exposed, E0, recovered, R0, and deceased, D0, are initialised to 0. The value of S0 is calculated by equation (14):

$$S0 = N-E0-I0-R0-D0 \qquad (14)$$

A study of variation in different parameters of the proposed Modified SEIRD model due to the Lockdowns imposed by the Indian government has also been included in the paper. By using the data of a Lockdown period, the value of β and other parameters have been calculated. There are four Lockdown and three Unlock periods in India till date. Hence, the data is divided in eight time periods: *Before Lockdown* [30th January – 24th March], *Lockdown 1.0* [25th March – 14th April], *Lockdown 2.0* [15th April – 3rd May], *Lockdown 3.0* [4th May – 17th May], *Lockdown 4.0* [18th May – 31st May], *Unlock 1.0* [1st June – 30th June], *Unlock 2.0* [1st July – 31st July] and *Unlock 3.0* [1st August onwards].

To obtain the optimal values of the parameters, the proposed Modified SEIRD model takes initial value, lower bound and upper bound as input for each parameter. For India, the initial value of β was taken as 0.5 with bounds [0.001, 1]. The parameter ε was initialised to 0.1 with bounds [0.0001, 1]. For α, the initial value was 1/5 and its bounds were [1/6, 1]. For parameter γ, the initial value was 0.1 with [1/14, 1] as bounds. For parameter $\mu$, 0.001 was the initial value and [0.001, 1] were the selected bounds. For the five states, the same initial values and bounds have been taken except for the lower bound of beta which was taken as 0.01. The optimal parameter values for curve fitting obtained for a particular Lockdown/Unlock period, are used for obtaining the projections in the next Lockdown/Unlock period.

The next sub-section describes the experimental setup required by the LSTM model for projecting the COVID-19 trend.

## 4.2. Experimental Setup for LSTM model

The six different time-series of Daily Confirmed, Daily Recovered, Daily Deceased, Total Confirmed, Total Recovered, and Total Deceased for India and its states obtained after pre-processing have been used for creating one LSTM for each of these series. These models have been created using the *keras* API. Since LSTM is very sensitive to range of data provided to it as input, therefore, pre-processing was required to normalize the input data to [0,1]. Normalization was done by using equation (15) through the *MinMaxScaler* function of *sklearn* library.

$$x_{norm} = \frac{x - x_{minimum}}{x_{maximum} - x_{minimum}} \qquad (15)$$

After normalization, the input data was split into 80-20 ratio as training and testing dataset respectively. Further, these datasets were divided into feature set and the output value set. The dimensionality of feature set and output value set depends on the hyper-parameter, *number_of_previous_days*. Its value signifies the number of previous days' output used to predict the output for the current day. Length of training set is taken as $n_{train}$ and value of *number_of_previous_days* as $p$. The dimensionality of feature set and output value set is taken as $(n_{train} - p \times p)$ and $(n_{train} - p \times 1)$ respectively.

The LSTM models designed for Confirmed and Recovered time-series, have four different LSTM layers and each layer has 100 LSTM cells. On the other hand, the LSTM model designed for Deceased time-series also has four LSTM layers with 150 LSTM cells per layer. After LSTM layers, one fully connected layer was added with 100 neurons in Confirmed and Recovered model and 150 neurons in Deceased model with linear activation function. After this, a fully connected layer was added with one neuron as output layer and linear activation function.

The LSTM model uses the Adam optimizer and Mean Squared Error (MSE) Loss with batch size 16 and 100 epochs in this paper. To avoid overfitting of the model, Dropout and Early Stopping was taken into account. In this paper, the probability of Dropout is taken as 0.4 and the patience value for Early Stopping as 5, i.e. if the testing loss has not improved in consecutive 5 epochs then the training will stop.

Finally, these models were fine-tuned against the hyper-parameter, *number_of_previous_days*, for the data of India and its five states having the highest number of total cases. The value of this hyper-parameter ranges between 1 and 90. The value of *number_of_previous_days* was selected with minimum testing loss. The optimal values of this hyper-parameter for India and its five states having the highest number of total cases, has been tabulated in Table 1.

Table 1: The optimal values of *number_of_previous_days* hyper-parameter

| State/Country | Optimal value of number_of_previous_days hyper-parameter | | | | | |
|---|---|---|---|---|---|---|
| | Daily Confirmed | Daily Recovered | Daily Deceased | Total Confirmed | Total Recovered | Total Deceased |
| India | 15 | 85 | 20 | 90 | 5 | 80 |
| Maharashtra | 50 | 30 | 20 | 20 | 65 | 35 |
| Tamil Nadu | 5 | 10 | 85 | 5 | 35 | 80 |
| Andhra Pradesh | 5 | 15 | 10 | 5 | 5 | 5 |
| Karnataka | 10 | 45 | 5 | 5 | 15 | 5 |
| Delhi | 75 | 35 | 25 | 5 | 50 | 90 |

The results obtained by both the proposed Modified SEIRD model and LSTM model are presented in the next section.

## 5. Results

This section discusses the results obtained by both the proposed Modified SEIRD model and LSTM models respectively. The proposed Modified SEIRD model performs both long-term and short-term projections, whereas, LSTM performs only short-term projections. Further, the 90% confidence intervals have been calculated using t-test. The corresponding results are discussed in detail in the following sub-sections.

## 5.1. Results obtained for Long-term predictions using the proposed Modifed SEIRD model

India is a geographically diverse nation, consisting of 36 states and union territories. Owing to India's large population and high population density, the Indian government implemented various nationwide lockdowns to curb the spread of Coronavirus. Though the COVID-19 pandemic did not leave any state/ union territory unaffected, some states have been far more affected by this deadly disease, as compared to others. As on August 15, 2020, the top five Indian states with the highest number of Total Confirmed cases are: Maharashtra, Tamil Nadu, Andhra Pradesh, Karnataka and Delhi. The curve for Daily Confirmed cases for India have been shown in Figure 5 (a), while Figure 5 (b) presents the curves for Exposed, Active Infected, Recovered, Deceased and Total Confirmed cases for India. Table 2 shows the parameter values obtained for India by the proposed Modified SEIRD model.

Table 2: Parameter values obtained for India

| India | Before Lockdown | Lockdown 1.0 | Lockdown 2.0 | Lockdown 3.0 | Lockdown 4.0 | Unlock 1.0 | Unlock 2.0 | Unlock 3.0 |
|---|---|---|---|---|---|---|---|---|
| $\beta$ | 0.161 | 0.306 | 0.095 | 0.239 | 0.084 | 0.081 | 0.119 | 0.088 |
| $\varepsilon$ | 0.999 | 0.384 | 0.0001 | 0.0001 | 0.0001 | 0.999 | 0.0001 | 0.0001 |
| $\alpha$ | 0.166 | 0.166 | 0.166 | 0.185 | 0.166 | 0.166 | 0.166 | 0.222 |
| $\gamma$ | 0.071 | 0.071 | 0.071 | 0.071 | 0.071 | 0.071 | 0.071 | 0.075 |
| $\mu$ | 0.007 | 0.007 | 0.008 | 0.001 | 0.006 | 0.003 | 0.002 | 0.001 |

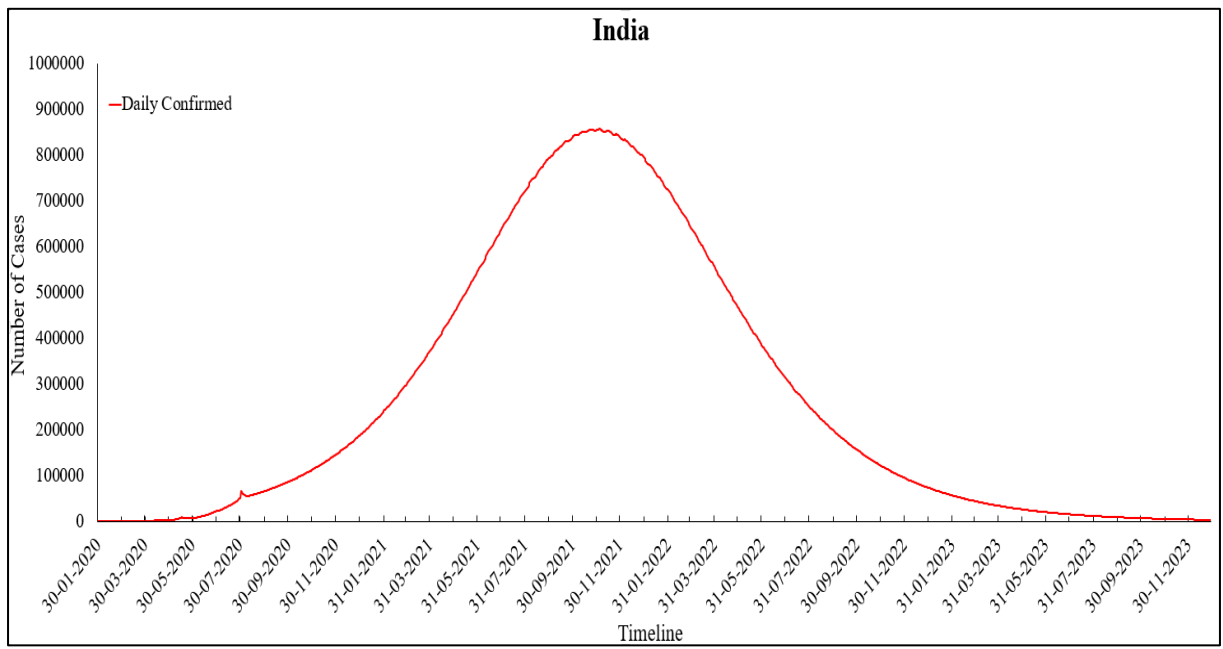

Figure 5 (a): Predicted Daily Confirmed cases for India

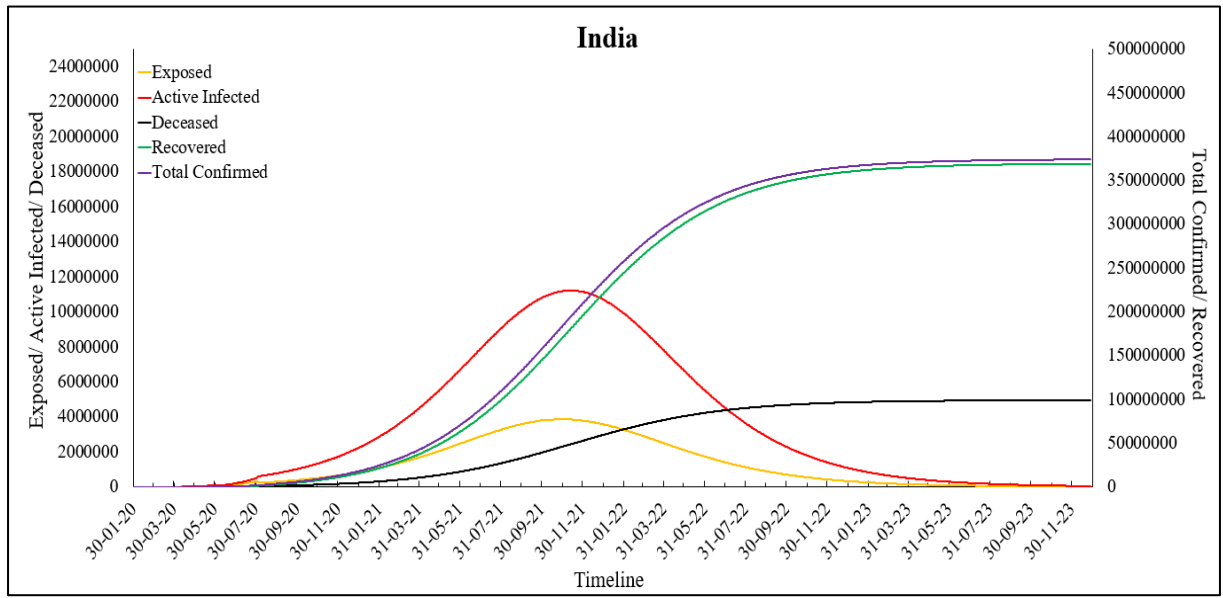

Figure 5 (b): Forecasting Results for India

The results obtained for India witness a peak of COVID-19 cases for India on 3rd November 2021 with 8,57,390 daily cases, and the disease may die out towards the end of year 2023. These results are based on the current situation; however, these may vary depending on various decisions taken by the Indian government and the precaution measures taken by people from time to time. It can be seen from Table 2 that the value of β, that represents the infection transmission rate per capita, increased during Lockdown 1.0. This resonates with the decision of the Central government to extend the lockdown. The effect of Lockdowns was also observed on ε, whose value reduced as lockdowns were extended and people became aware about the importance of precautionary measures. But during Unlock 1.0, the value of ε increased, thus causing a rise in the number of exposed people. It was observed that both the parameters ε and µ decreased over time, except during Unlock1.0 when many relaxations were given by the government, more people were exposed and the value of ε suddenly increased. The increase in the value of α signifies a reduction in the average latent period and an increased γ depicts a reduction in the time to recover $1/\gamma$, resulting in increase in recovery rate among the Indian population. There has also been a reduction in the value of reproduction number for India, and it has reduced from $R_0 = 3.028$ to $R_e = 1.163$ as on 15th August 2020. This indicates that there has been a reduction in the number of secondary infections produced by an infected person. This positive change in parameter values can be attributed to the restrictions imposed by the Indian authorities. Next, Table 3 presents the parameter values obtained for the Maharashtra state, while its forecasting results have been shown in Figure 6 (a) and Figure 6 (b).

Table 3: Parameter values obtained for Maharashtra

| MH | Before Lockdown | Lockdown 1.0 | Lockdown 2.0 | Lockdown 3.0 | Lockdown 4.0 | Unlock 1.0 | Unlock 2.0 | Unlock 3.0 |
|---|---|---|---|---|---|---|---|---|
| β | 0.297 | 0.239 | 0.092 | 0.075 | 0.071 | 0.062 | 0.067 | 0.087 |
| ε | 0.001 | 0.001 | 0.999 | 0.999 | 0.999 | 0.999 | 0.999 | 0.0001 |
| α | 0.999 | 0.434 | 0.397 | 0.515 | 0.999 | 0.499 | 0.977 | 0.279 |
| γ | 0.010 | 0.019 | 0.018 | 0.023 | 0.043 | 0.041 | 0.045 | 0.074 |
| µ | 0.003 | 0.013 | 0.003 | 0.003 | 0.002 | 0.004 | 0.002 | 0.002 |

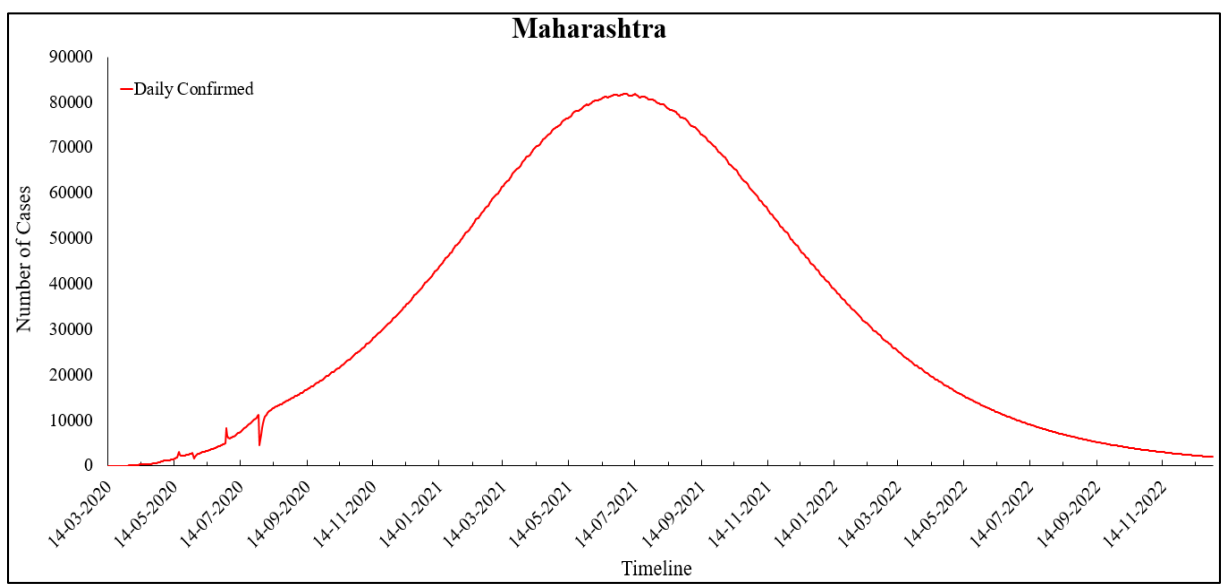

Figure 6 (a): Predicted Daily Confirmed cases for Maharashtra

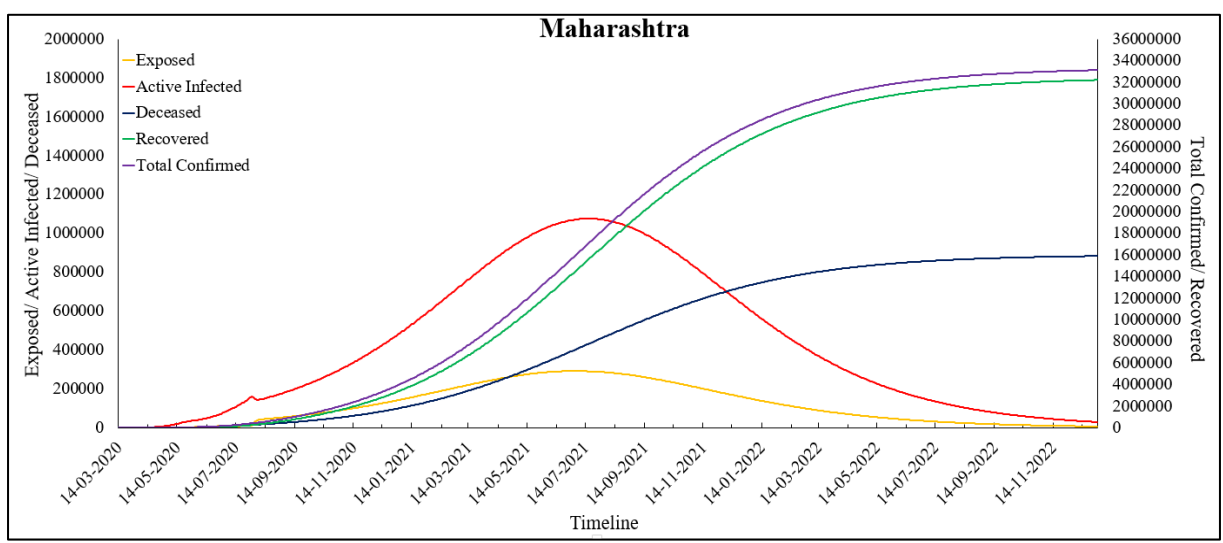

Figure 6 (b): Forecasting Results for Maharashtra

The state of Maharashtra has witnessed the highest number of Coronavirus cases since the advent of the pandemic in India. The main reason behind this can be Maharashtra's high population density and the presence of Asia's largest slum named Dharavi. Due to this, Maharashtra had a very high Ɍ0 in the initial days of the pandemic. However, by mid-August 2020, Ɍe has reduced below 1.2 as computed by the proposed model. The proposed Modified SEIRD model indicates that on 5$^{th}$ July 2021, Maharashtra will record 81,935 Daily Confirmed cases which would be the highest for this state. The number of Total Confirmed cases would have crossed 16 million by that time, out of which 14 million would have recovered from this disease. For Maharashtra, the death rate μ has always remained higher than most of the other Indian states, and the recovery rate γ has been lower than the national average. Due to the severity of the infection, the Maharashtra government decided to extend the duration of restrictions imposed on its residents. However, several relaxations were offered during Unlock 3.0 and the impact of this has been captured by the β parameter of the proposed model. Its value sees a decreasing trend till Unlock 2.0, but an increase in this value can be seen during Unlock 3.0. Surprisingly, a downfall in the proportion of Exposed population is observed at the same time. However, under the prevailing circumstances, the proposed model predicts more than 0.8 million COVID-19 deaths in the state by the end of year 2022. Table 4 shows the parameter values obtained for Tamil Nadu, while its forecasting results for Tamil Nadu have been shown in Figure 7 (a) and Figure 7 (b).

Table 4: Parameter values obtained for Tamil Nadu

| TN | Before Lockdown | Lockdown 1.0 | Lockdown 2.0 | Lockdown 3.0 | Lockdown 4.0 | Unlock 1.0 | Unlock 2.0 | Unlock 3.0 |
|---|---|---|---|---|---|---|---|---|
| β | 0.451 | 0.207 | 0.001 | 0.339 | 0.045 | 0.152 | 0.103 | 0.078 |
| ε | 0.999 | 0.999 | 0.0001 | 0.0001 | 0.999 | 0.0001 | 0.0001 | 0.999 |
| α | 0.166 | 0.293 | 0.166 | 0.999 | 0.166 | 0.282 | 0.216 | 0.195 |
| γ | 0.071 | 0.071 | 0.071 | 0.071 | 0.071 | 0.071 | 0.080 | 0.124 |
| μ | 0.016 | 0.002 | 0.012 | 0.001 | 0.002 | 0.001 | 0.001 | 0.002 |

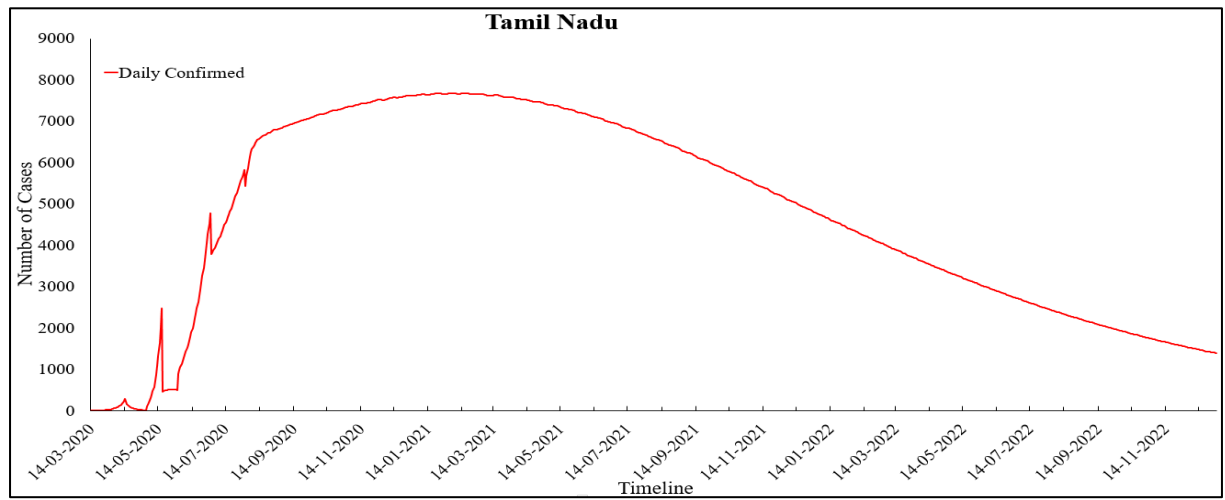

Figure 7 (a): Predicted Daily Confirmed cases for Tamil Nadu

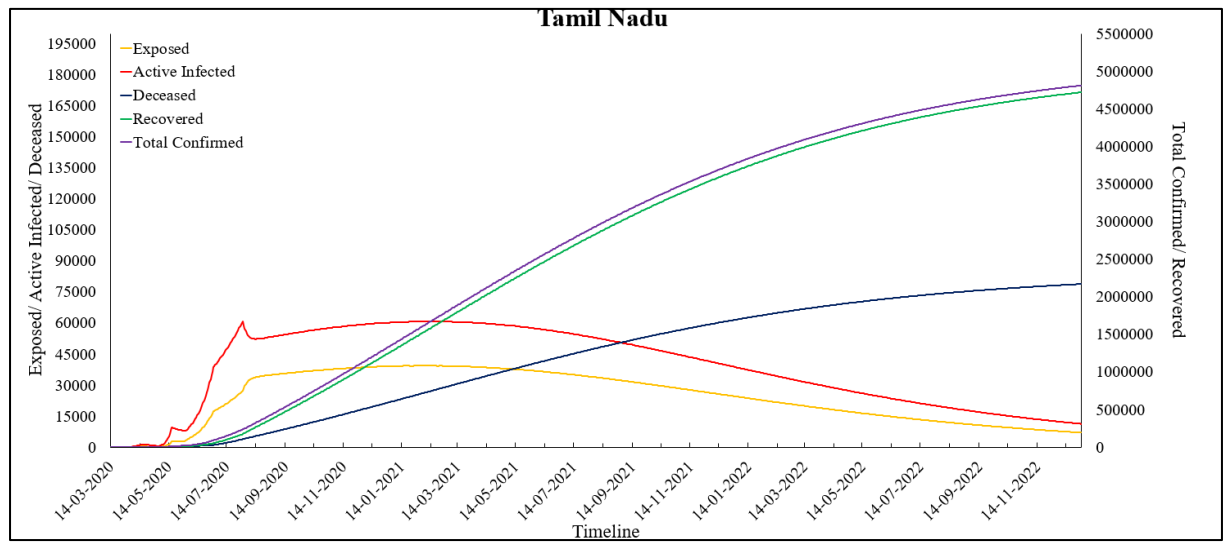

Figure 7 (b): Forecasting Results for Tamil Nadu

As of 15th August 2020, Tamil Nadu had the second highest number of COVID-19 cases in India and has crossed 0.3 million Total Confirmed cases. The values of Ɍ0 and Ɍe for this state were found to be 7.86 and 1.02 (as on 15th August 2020) respectively, as calculated by the proposed model. Out of these, almost 80% people have recovered and 1.5% people have succumbed to this deadly disease. This southern state of India witnessed a sharp spike in the number of COVID-19 cases since June 2020 and based on the forecasting results of the proposed Modified SEIRD model, Tamil Nadu will have the highest Daily Confirmed cases on 5th February 2021 with a count of 7,681. By the end of year 2022, it is projected that 6% of the state's current population would have been infected from the virus and around 0.1% people would lose their battle against this pandemic. As shown in Table 4, the parameter values computed from the model, indicate a hike in the recovery rate ($\gamma$) as well as the death rate ($\mu$) in the state. Next, Table 5 presents the parametric values obtained by the proposed model for Andhra Pradesh, while Figure 8 (a) and Figure 8 (b) represent its forecasting results.

Table 5: Parameter values obtained for Andhra Pradesh

| AP | Before Lockdown | Lockdown 1.0 | Lockdown 2.0 | Lockdown 3.0 | Lockdown 4.0 | Unlock 1.0 | Unlock 2.0 | Unlock 3.0 |
|---|---|---|---|---|---|---|---|---|
| β | 0.316 | 0.241 | 0.095 | 0.001 | 0.122 | 0.166 | 0.122 | 0.096 |
| ε | 0.248 | 0.999 | 0.999 | 0.0001 | 0.999 | 0.0001 | 0.999 | 0.0001 |
| α | 0.477 | 0.999 | 0.166 | 0.166 | 0.166 | 0.166 | 0.166 | 0.166 |
| γ | 0.071 | 0.071 | 0.071 | 0.071 | 0.071 | 0.071 | 0.071 | 0.096 |
| μ | 0.001 | 0.003 | 0.009 | 0.001 | 0.001 | 0.002 | 0.002 | 0.001 |

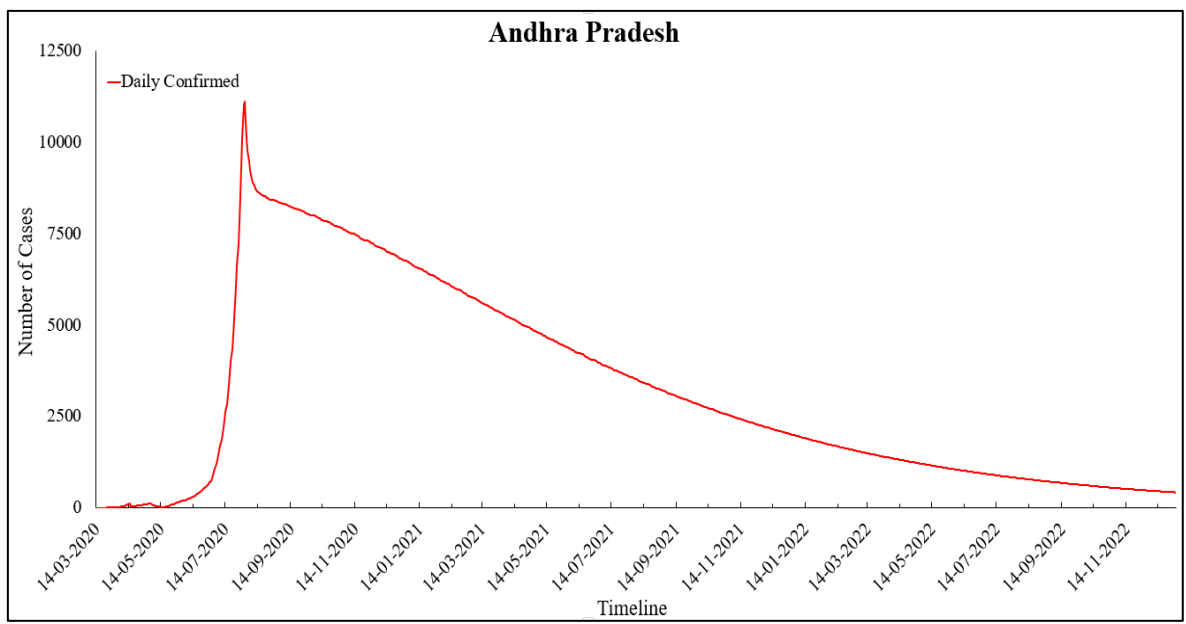

Figure 8 (a): Predicted Daily Confirmed cases for Andhra Pradesh

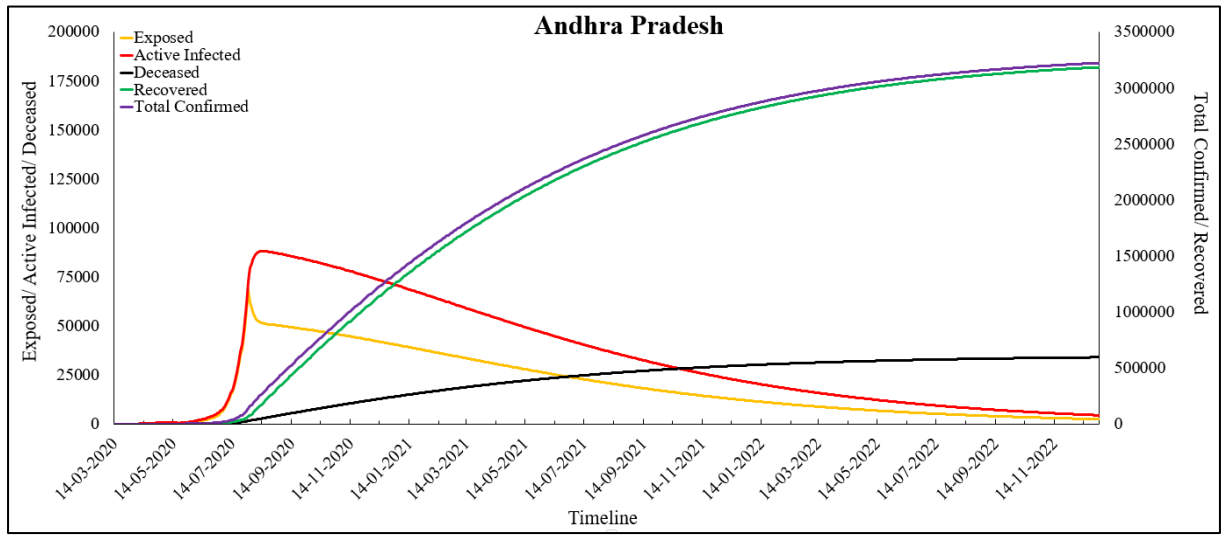

Figure 8 (b): Forecasting Results for Andhra Pradesh

For Andhra Pradesh, it can be seen that the per capita transmission rate of the disease, represented by β, significantly increased during the fourth Lockdown and the first Unlock period. However, at the time of writing of this paper Andhra Pradesh is the only state that has surpassed its COVID-19 peak. The state observed its peak on the first day of August 2020 with the maximum recorded Daily Confirmed cases of 11,122. It can also be verified from Figure 8 (a). There is a significant difference in the shape of the curve obtained for Andhra Pradesh and Tamil Nadu. The main reason behind this is that both these states saw a steep rise in the number of COVID-19 cases around the second half of the year 2020. This scenario can also be seen by observing the trend of their parameter values. The infection curve for Andhra Pradesh has started seeing a downward trend, which is also reflected by the increasing value of the recovery rate. Also, the value of reproduction number for Andhra Pradesh has gone below the value 1, which clearly indicates that the decline of this deadly disease in the state. Next, Table 6 presents the parameter values for the state of Karnataka, while Figure 9 (a) and Figure 9 (b) present the forecasting results for the same.

Table 6: Parameter values obtained for Karnataka

| KA | Before Lockdown | Lockdown 1.0 | Lockdown 2.0 | Lockdown 3.0 | Lockdown 4.0 | Unlock 1.0 | Unlock 2.0 | Unlock 3.0 |
|---|---|---|---|---|---|---|---|---|
| β | 0.252 | 0.091 | 0.068 | 0.142 | 0.085 | 0.104 | 0.157 | 0.062 |
| ε | 0.816 | 0.999 | 0.999 | 0.248 | 0.999 | 0.047 | 0.0001 | 0.999 |
| α | 0.518 | 0.999 | 0.999 | 0.166 | 0.387 | 0.742 | 0.999 | 0.166 |
| γ | 0.015 | 0.023 | 0.043 | 0.046 | 0.026 | 0.071 | 0.049 | 0.064 |
| μ | 0.010 | 0.003 | 0.003 | 0.002 | 0.001 | 0.001 | 0.003 | 0.001 |

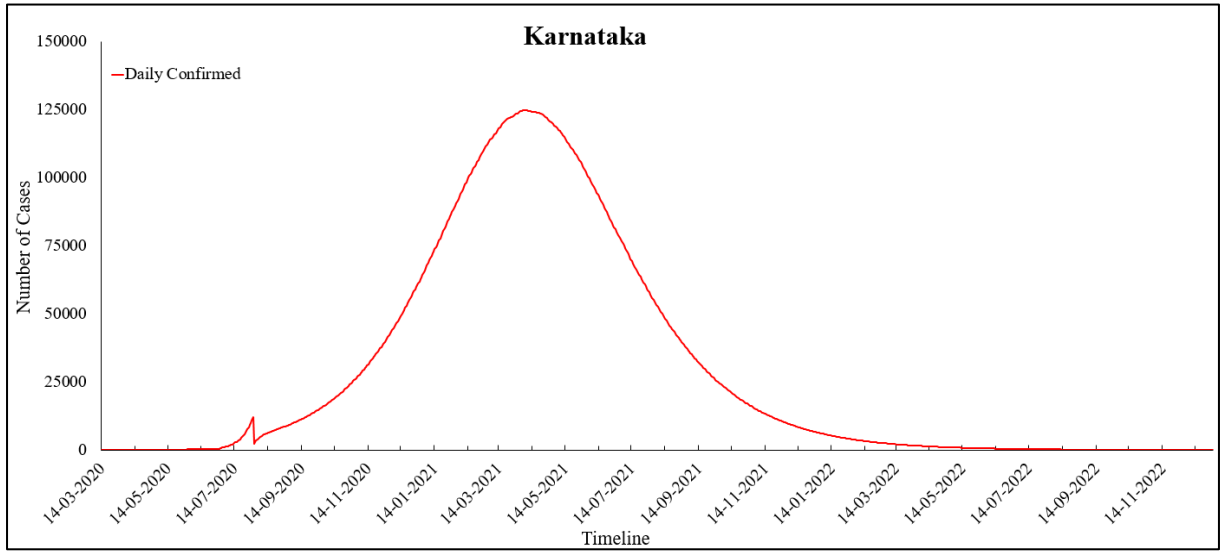

Figure 9 (a): Predicted Daily Confirmed cases for Karnataka

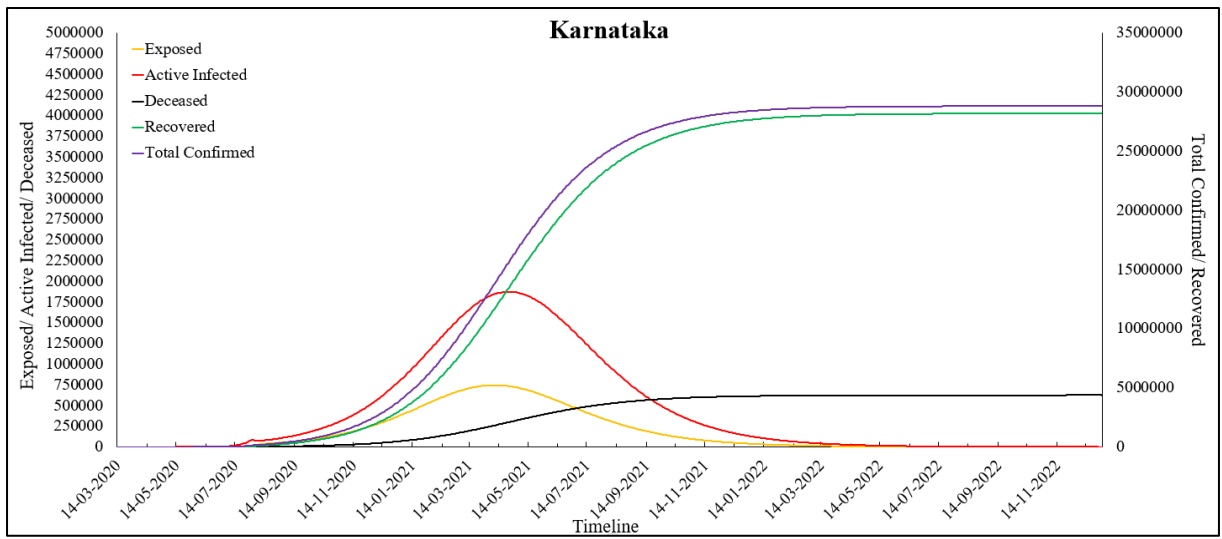

Figure 9 (b): Forecasting Results for Karnataka

The results obtained for Karnataka show that there will be a peak of COVID-19 cases on 7th April 2021 with 1,24,722 Daily infections, and the disease will die out towards the end of year 2022. The values of Ʀ0 and Ʀe for Karnataka were found to be 9.93 and 1.31 (as on 15th August 2020) respectively, as calculated by the proposed model. An interesting trend was observed in the values of β and ε. When one of these parameters decreases, the other increases and vice-versa. Overall, β has reduced over time, which depicts a reduction in the transmission rate per capita in this state. It was seen that during the first two Unlocks, ε also reduced. However, Unlock 3.0 witnessed a surge in its value, thereby indicating an increase in the number of exposed individuals as more relaxations were offered to the public. On an average, it was found that γ, the recovery rate has increased in the state, leading to a higher number of recovered individuals. It can be seen from the results that the value of μ, that denotes the rate of deceased population, has decreased over time. Table 7 shows the parameter values for Delhi and its forecasting results have been shown in Figure 10 (a) and Figure 10 (b).

Table 7: Parameter values obtained for Delhi

| DL | Before Lockdown | Lockdown 1.0 | Lockdown 2.0 | Lockdown 3.0 | Lockdown 4.0 | Unlock 1.0 | Unlock 2.0 | Unlock 3.0 |
|---|---|---|---|---|---|---|---|---|
| β | 0.235 | 0.274 | 0.038 | 0.157 | 0.111 | 0.083 | 0.037 | 0.116 |
| ε | 0.991 | 0.806 | 0.0001 | 0.999 | 0.999 | 0.999 | 0.0001 | 0.999 |
| α | 0.516 | 0.282 | 0.166 | 0.999 | 0.166 | 0.166 | 0.166 | 0.999 |
| γ | 0.071 | 0.071 | 0.071 | 0.071 | 0.071 | 0.071 | 0.106 | 0.119 |
| μ | 0.013 | 0.004 | 0.009 | 0.001 | 0.005 | 0.003 | 0.002 | 0.001 |

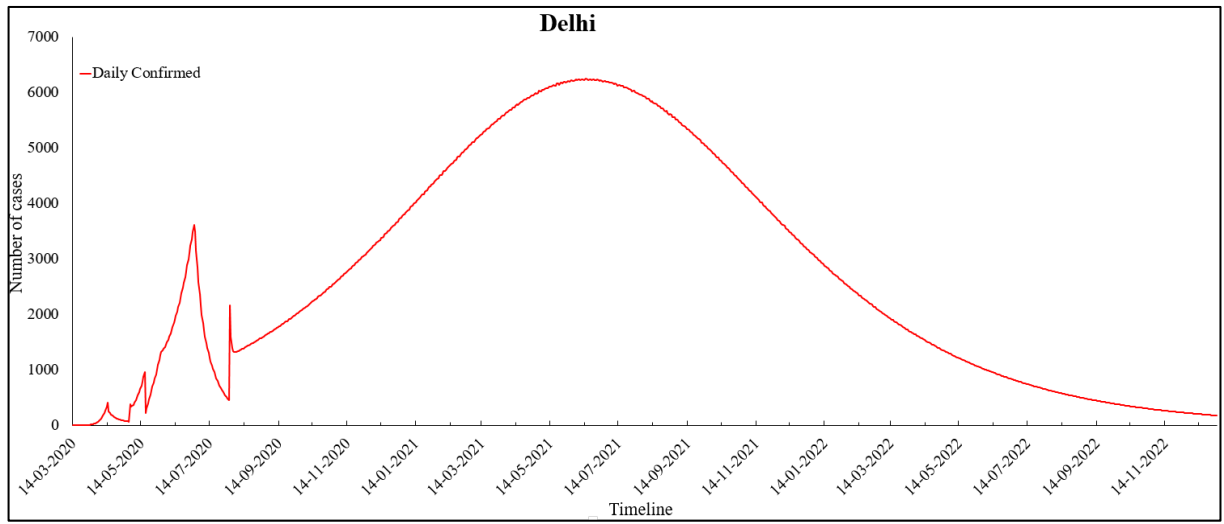

Figure 10 (a): Predicted Daily Confirmed cases for Delhi

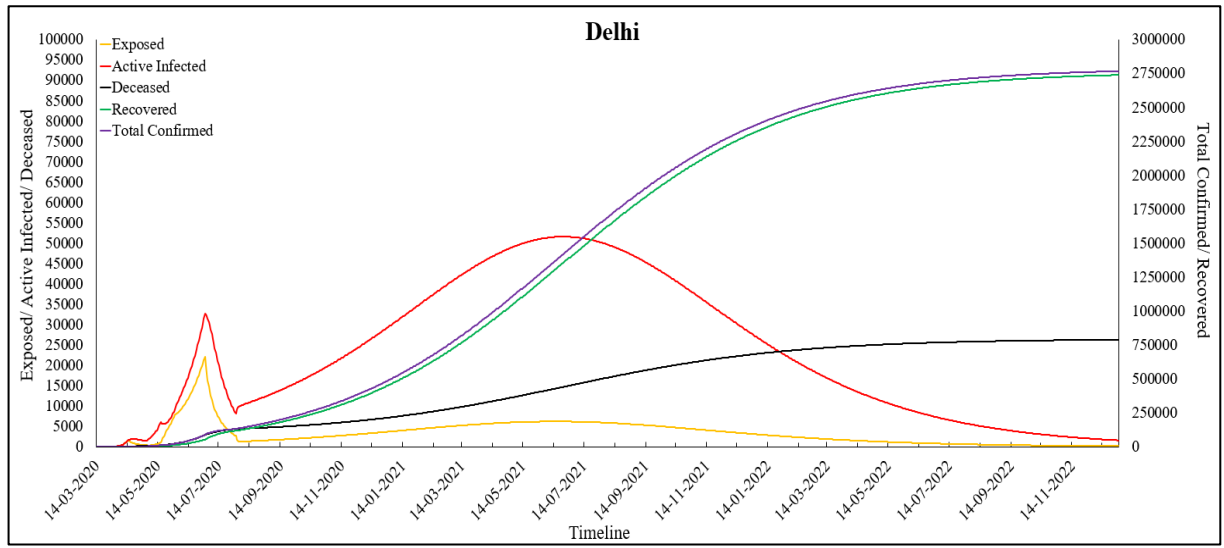

Figure 10 (b): Forecasting Results for Delhi

Delhi is the national capital of India which has always remained in the top five states till 15[th] August, and has been severely affected by COVID-19. However, Delhi's rank has drastically changed over the recent months and it has now shifted down from the second spot to the fifth spot. The value of reproduction number for Delhi has also come down from 3.21 to 1.06 as of 15th August 2020. Despite this fact, the projections for Delhi indicate that the current scenario is just a local-minima and its infection curve will reach its peak on 15th June 2021 with 6249 Daily Confirmed cases. At the time of peak, the number of Daily Infected cases is expected to rise to 6249, with more than 51k Active Infected people of Delhi. According to the proposed Modified SEIRD model, Delhi will have more than 2 million Total cases by the end of 2022, out of which 98% people would have recovered. This attributes to the high recovery rate of this union territory. However, the sharp decrease in the number of cases in Delhi, has also brought the authorities under a scanner. Multiple news channels have questioned the authenticity of the quality of the tests being conducted for COVID-19 patients.

## 5.2. Results obtained for Short-term prediction using the proposed Modified SEIRD model and LSTM model

This section discusses the short-term projections performed using the proposed Modified SEIRD model as well as the LSTM model for India and its five worst-affected states for next 30 days, i.e., from 16th August 2020 till 14th September 2020. The predictions for six time-series namely, Daily Confirmed, Daily Recovered, Daily Deceased, Total Confirmed, Total Recovered and Total Deceased, have been obtained from both the models. Case Fatality Rate, Recovery Rate and Mean Absolute Percentage Error (MAPE) have also been calculated to analyse the results. These have been mathematically defined in equation (16), (17) and (18) respectively.

$$\text{Case Fatality Rate} = \frac{\text{Total Deceased}}{\text{Total Confirmed}} \times 100 \tag{16}$$

$$\text{Recovery Rate} = \frac{\text{Total Recovered}}{\text{Total Confirmed}} \times 100 \tag{17}$$

$$\text{Mean Absolute Percentage Error} = \frac{1}{n} * \sum_{1}^{n} \frac{|\text{Actual value} - \text{Forecsted value}|}{\text{Actual Value}} \tag{18}$$

The Recovery Rate and Case Fatality Rate on 15th August 2020 and on 14th September 2020 (30 days from now) for both the models, have been given in Table 8 and Table 9 respectively.

Table 8: Recovery and Case Fatality Rate for India and its five worst affected states on 15th August 2020

| Country/ State | Actual Recovery Rate (%) | SEIRD Recovery Rate (%) | LSTM Recovery Rate (%) | Actual Case Fatality Rate (%) | SEIRD Case Fatality Rate (%) | LSTM Case Fatality Rate (%) |
|---|---|---|---|---|---|---|
| India | 71.86 | 71.55 | 68.87 | 1.93 | 2.08 | 1.84 |
| Maharashtra | 69.82 | 70.50 | 70.50 | 3.38 | 3.34 | 2.34 |
| Tamil Nadu | 81.98 | 82.67 | 80.92 | 1.70 | 1.67 | 1.66 |
| Andhra Pradesh | 67.82 | 67.48 | 72.75 | 0.91 | 1.08 | 0.95 |
| Karnataka | 61.30 | 59.66 | 62.27 | 1.75 | 1.97 | 1.88 |
| Delhi | 89.68 | 89.83 | 100 | 2.76 | 2.97 | 2.79 |

Table 9: Predicted Recovery and Case Fatality Rate for India and its five worst affected states on 14[th] September 2020

| Country/ State | SEIRD Recovery Rate (%) | LSTM Recovery Rate (%) | SEIRD Case Fatality Rate (%) | LSTM Case Fatality Rate (%) |
|---|---|---|---|---|
| India | 72.26 | 69.25 | 2.04 | 1.81 |
| Maharashtra | 79.78 | 46.82 | 2.80 | 1.20 |
| Tamil Nadu | 89.80 | 74.47 | 1.64 | 1.64 |
| Andhra Pradesh | 86.29 | 100 | 1.06 | 1.05 |
| Karnataka | 71.84 | 85.46 | 1.79 | 2.33 |
| Delhi | 90.95 | 90.29 | 2.23 | 2.70 |

It can be observed from Table 8 that the values of Recovery Rate calculated by both the models are very close to the Actual Recovery Rate values on 15[th] August 2020. Similarly, the Case Fatality Rates computed by both the models for India and its five states in consideration, almost resemble the Actual Case Fatality Rates. Moreover, the Recovery Rates and Case Fatality Rates achieved by the proposed model closely correspond the values obtained by LSTM as well. This clearly indicates that the proposed Modified SEIRD model can accurately forecast the trend of COVID-19 in India and its states by modelling all the important parameters correctly.

The short-term projections performed by the proposed Modified SEIRD model show that by 14[th] September 2020, the Recovery Rate for India may increase by more than 1% and the Case Fatality Rate may decrease by less than 0.5%. For the state of Maharashtra, the proposed model shows an improvement of around 9% in the Recovery Rate and a drop of about 0.5% in the Case Fatality Rate. Surprisingly, LSTM shows a drastic reduction in the value of Recovery Rate for this state. A similar pattern can be observed for Tamil Nadu, for which the proposed model predicts a positive change in the Recovery Rate, but the LSTM model predicts a reduction in this value. However, both the models agree on a gradual decline in the value of Case Fatality Rate for Tamil Nadu.

For the remaining three states, the proposed Modified SEIRD model and the LSTM have similar projections for the Recovery Rate of COVID-19 disease in upcoming 30 days. Regardless, a contrasting trend in the values of Case Fatality Rates can be observed for Andhra Pradesh and Karnataka, where one model predicts a hike but the other indicates a decline in the value. Figure 11 (a-f) present the projections made by both the models for Daily Confirmed cases in India and its five worst affected states. The lower and upper estimates corresponding to 90% confidence interval (CI) has also been shown in the graphs for the predictions made by both the models.

Based on the projections of the proposed Modified SEIRD model, by mid-September, the number of Daily Confirmed in India, Maharashtra and Tamil Nadu will be 74733; 16906 and 6947 respectively. For Andhra Pradesh, Karnataka and Delhi, this number is expected to be 8240; 11357 and 1787 respectively. Similarly, by using the LSTM model the estimated number of Daily Confirmed cases are 82190; 14750; 6266; 11311; 10274 and 1067 for India, Maharashtra, Tamil Nadu, Andhra Pradesh, Karnataka and Delhi.

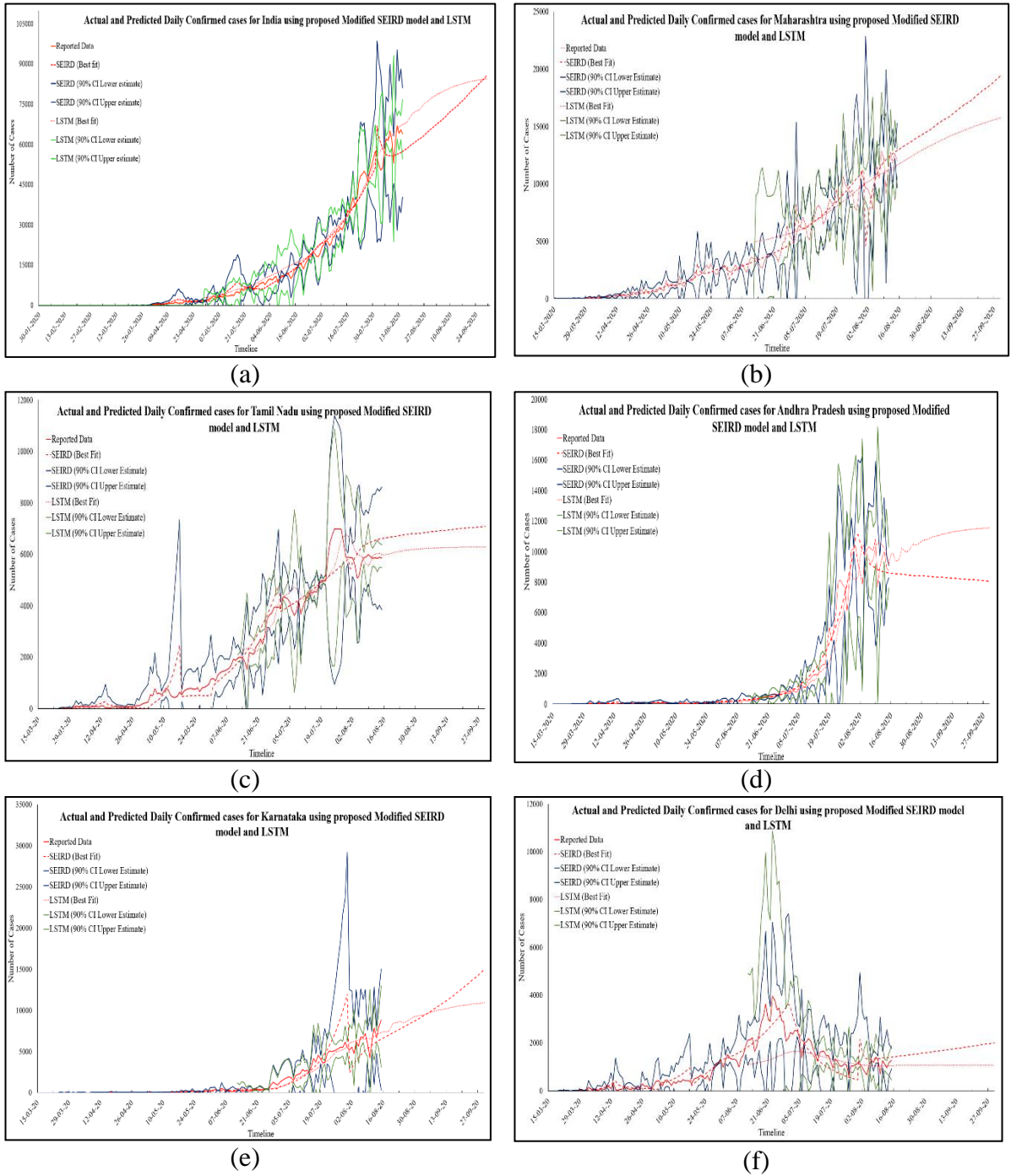

Figure 11 (a-f): Actual and Predicted Daily Confirmed cases for India and its five worst-affected states

As is evident from Figure 11, that the predictions made by both the models were found to be significantly similar to each other. This claim is also verified by the values of Mean Absolute Percentage Error (MAPE) for both the models. For the proposed Modified SEIRD model, these values were 0.05, 0.11, 0.11, 0.15, 0.33 and 0.29 for India and its five states respectively. In case of LSTM model, MAPE values were 0.18, 0.09, 0.18, 0.28 and 0.28 for India, Maharashtra, Tamil Nadu, Andhra Pradesh, Karnataka and Delhi respectively. Figure 12 (a-f) present the projections made by both the models for Daily Recovered cases in India and its five worst affected states along with their 90% confidence intervals obtained through t-test.

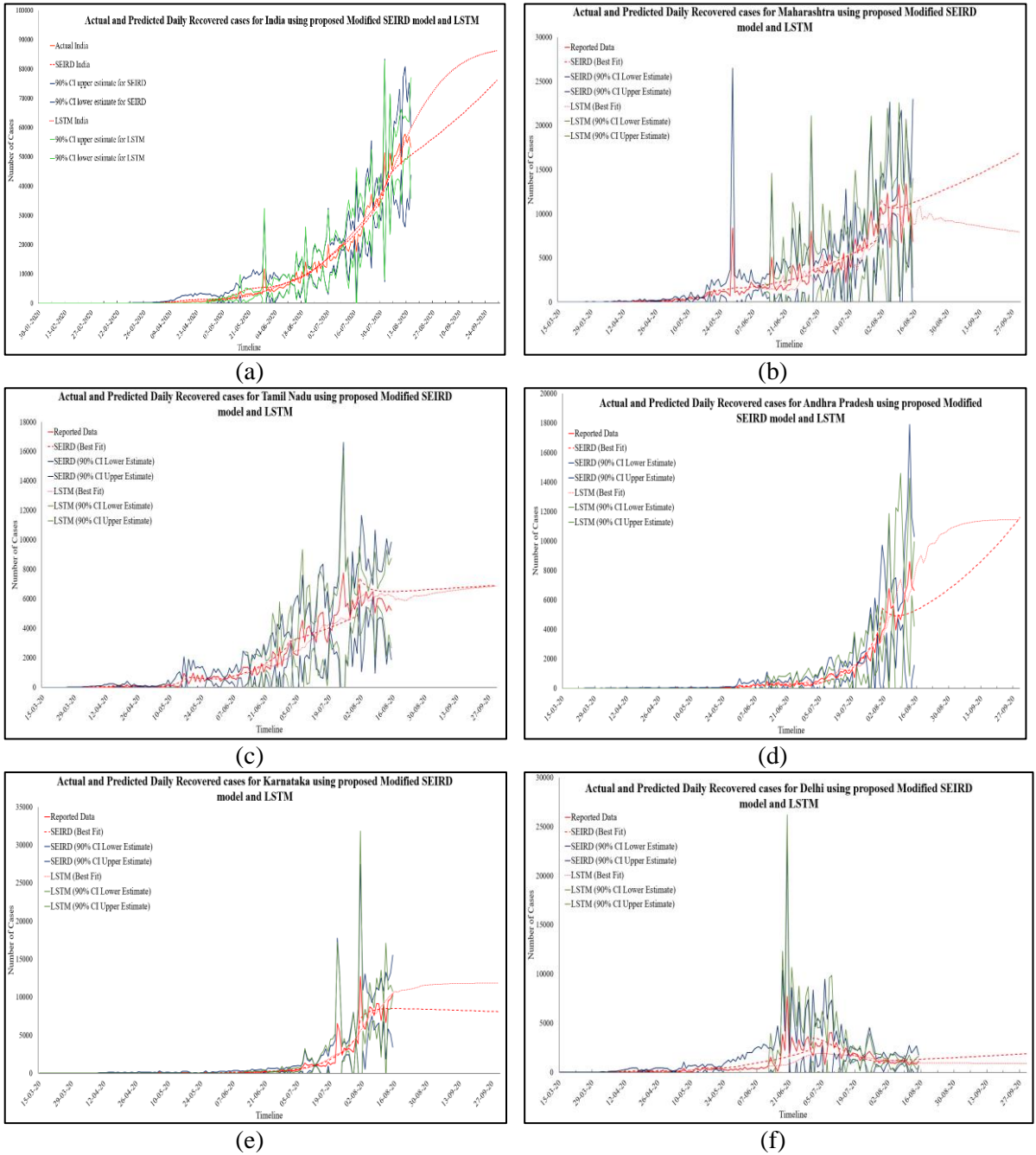

Figure 12 (a-f): Actual and Predicted Daily Recovered cases for India and its five worst-affected states

The graphs in Figure 12 show a rising trend in the number of Daily Recovered cases. According to the proposed Modified SEIRD model, there will 66119; 14703; 6766; 8263; 8762 and 1665 Daily Recovered cases in India, Maharashtra, Tamil Nadu, Andhra Pradesh, Karnataka and Delhi by 14[th] September 2020 respectively. However, based on the forecasting performed by the LSTM model, India and its five states will have 82891; 8547; 6624; 11805; 11371 and 913 Daily Recovered cases. The proposed model was able to perform accurate predictions for this Daily Recovered time series data. The MAPE values obtained from the proposed model were 0.06, 0.25, 0.21, 0.36, 0.30 and 0.59 for India, Maharashtra, Tamil Nadu, Andhra Pradesh, Karnataka and Delhi respectively. Figure 13 (a-f) present the projections made by both the models for Daily Deceased cases in India and its five worst affected states along with the lower and upper estimates corresponding to 90% confidence intervals.

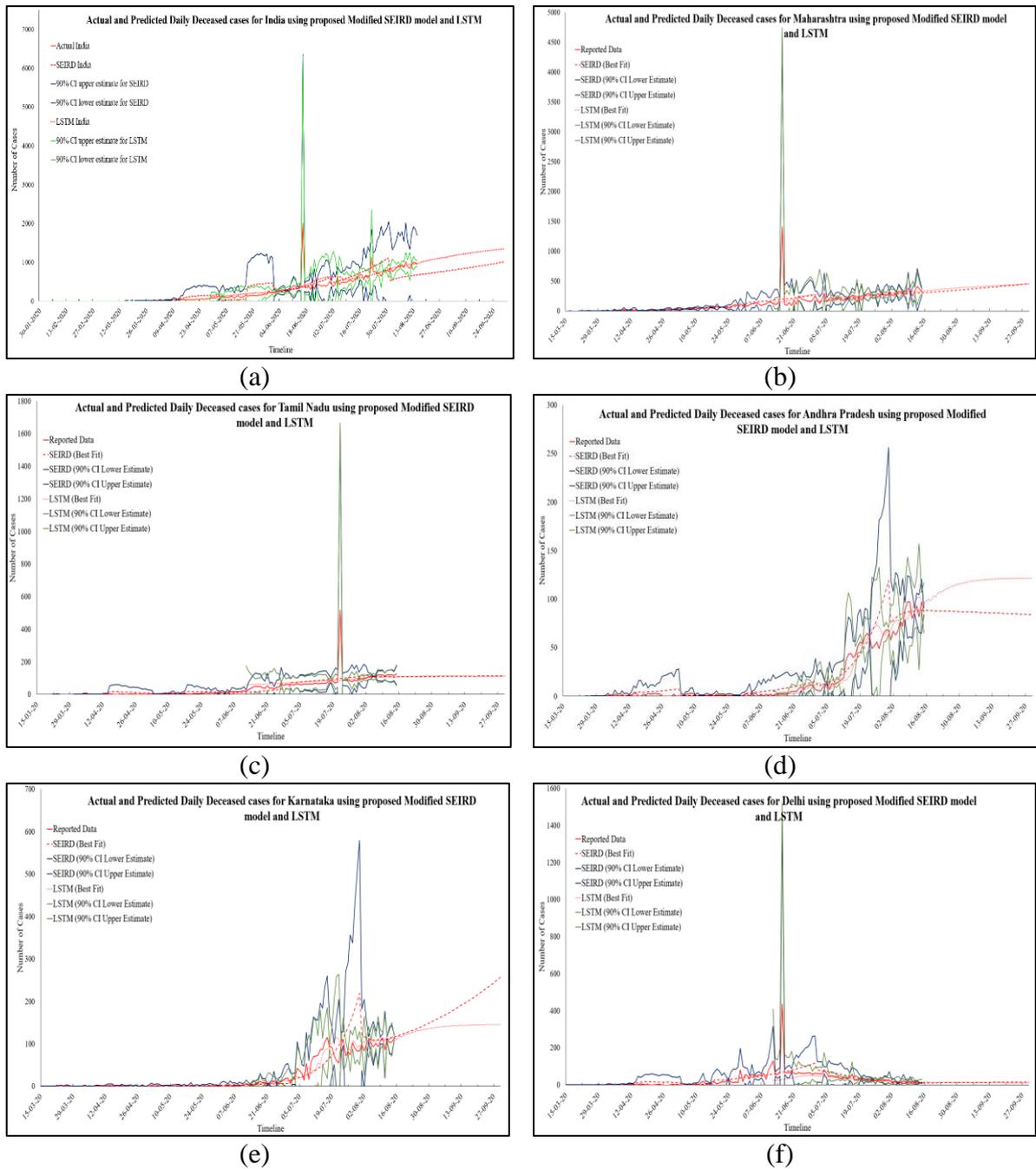

Figure 13 (a-f): Actual and Predicted Daily Deceased cases for India and its five worst-affected states

To lessen the impact of any pandemic, it is important to control the number of deceased patients. The trend of Daily Deceased individuals for next 30 days can be observed in Figure 13 (a-f). The proposed Modified SEIRD model and the LSTM model predict 881 and 1254 Daily Deceased for India on 14$^{th}$ September 2020. For the five Indian states in consideration, this value will remain below five hundred, with 400, 112, 86, 194 and 14 Daily Death cases estimated by the proposed model; and 423, 112, 121, 144 and 12 Daily Deceased cases as per the LSTM model. Using the Daily Deceased cases, the proposed Modified SEIRD model was able to obtain extremely accurate predictions. The MAPE values corresponding to both the models, was found to be below 0.60 for India and its five states in consideration. Figure 14 (a-

f) present the projections made by both the models for Total Confirmed cases in India and its five worst affected states along with their confidence intervals obtained through t-test.

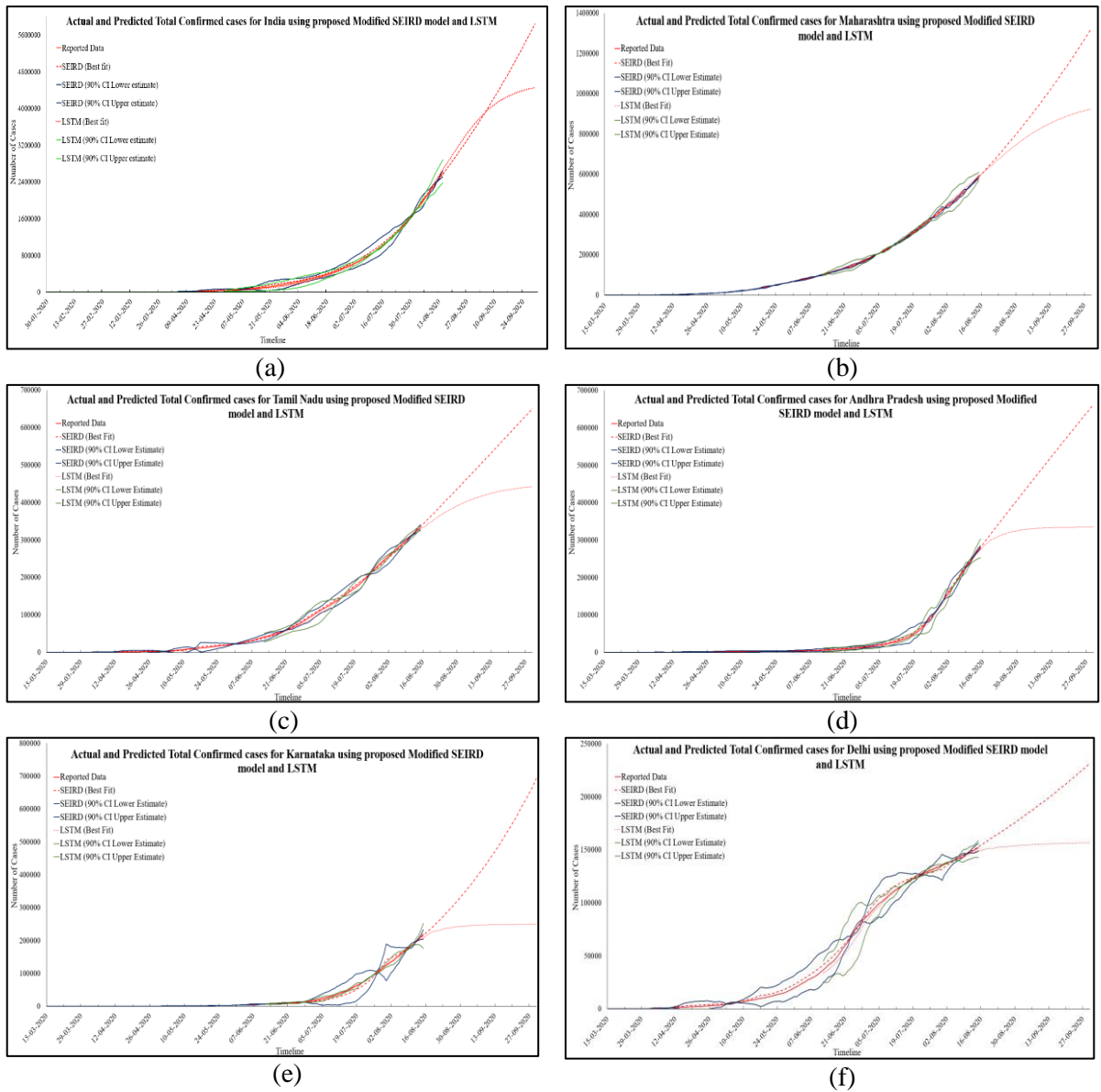

Figure 14 (a-f): Actual and Predicted Total Confirmed cases for India and its five worst-affected states

The forecasting results of the proposed Modified SEIRD model show that the number of Total Confirmed in India, Maharashtra and Tamil Nadu will be 4547801; 1034009 and 538699 respectively. For Andhra Pradesh, Karnataka and Delhi, this number is expected to be 532542; 481769 and 201145 respectively. Similarly, by using the LSTM model the estimated number of Total Confirmed cases are 4200226; 612359; 425972; 333376; 248125 and 155748 for India, Maharashtra, Tamil Nadu, Andhra Pradesh, Karnataka and Delhi. For the same time-series, the MAPE values for the proposed model were found to be 0.02, 0.01, 0.02, 0.08, 0.1 and 0.01 corresponding to India and its five states respectively. For the LSTM model, the corresponding MAPE values were 0.01, 0.02, 0.03, 0.09, 0.04 and 0.03. These low error values indicate the effectiveness of the proposed Modified SEIRD model in performing projections. Figure 15 (a-f) present the projections made by both the models for Total Recovered cases in India and its five worst affected states along with their confidence intervals obtained through t-test.

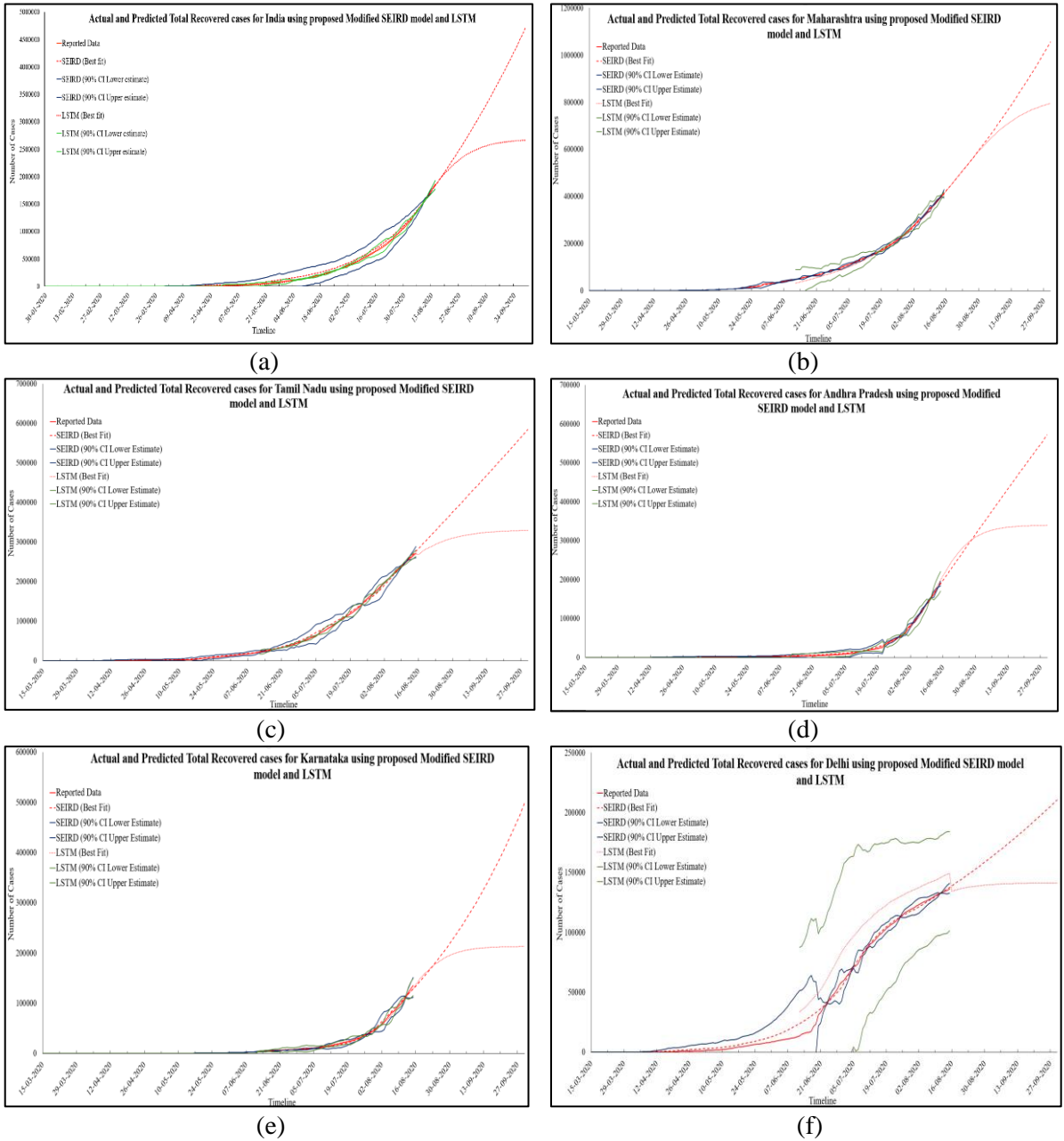

Figure 15 (a-f): Actual and Predicted Total Recovered cases for India and its five worst-affected states

Figure 15 depicts an increasing number of Total Recovered cases in the upcoming 30 days. According to the projections made by the proposed Modified SEIRD model, there will be 3585678; 803056; 475286; 441233; 334757 and 182244 Total Recovered cases in India, Maharashtra, Tamil Nadu, Andhra Pradesh, Karnataka and Delhi by 14$^{th}$ September 2020 respectively. However, based on the predictions obtained from the LSTM model, India and its five states will have 2595423; 318204; 324921; 335020; 210364 and 140954 Total Recovered cases. Based on the above values, the MAPE for India was found to be 0.07 and 0.01 for the proposed Modified SEIRD model and LSTM model respectively. For Maharashtra, Tamil Nadu, Andhra Pradesh, Karnataka and Delhi, the proposed model achieved MAPE of 0.01, 0.05, 0.2, 0.05 and 0.1 respectively; whereas the LSTM model had MAPE values of 0.08, 0.01, 0.18, 0.09 and 0.18 respectively. Figure 16 (a-f) present the projections made by both the models for Total Deceased cases in India and its five worst affected states along with their confidence intervals obtained through t-test.

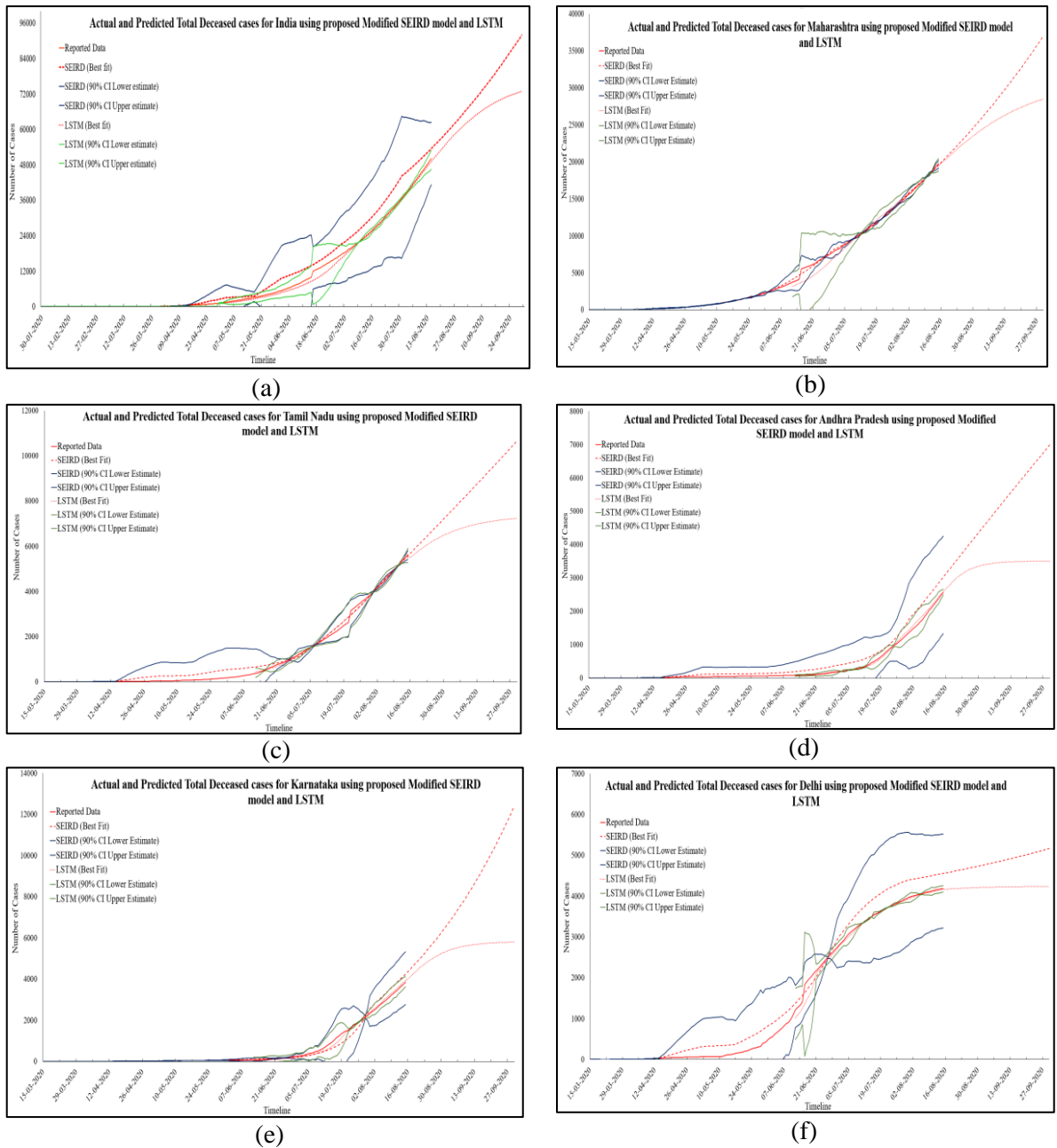

Figure 16 (a-f): Actual and Predicted Total Deceased cases for India and its five worst-affected states

The proposed Modified SEIRD model and the LSTM model predict 76724 and 68132 Total Deceased for India on 14$^{th}$ September 2020. For the five Indian states in consideration, this count will be 30163; 8858; 5639; 8811 and 4927 Total Death cases estimated by the proposed model; and 8517; 7003; 3487; 5687 and 4228 Total Deceased cases as per the LSTM model. For this time series data, it was found that the projections of the proposed model were highly synced with the projections made by LSTM. This can be verified from the fact that for India, the proposed model and LSTM achieved low MAPE values of 0.11 and 0.03 respectively. Furthermore, the MAPE values for the proposed Modified SEIRD model was found to be below 0.8 for all the five states in consideration.

# 6. Conclusion

With the exponential growth in the number of Corona Virus Disease 2019 (COVID-19) cases worldwide, countries need to prepare themselves with appropriate measures to tackle this epidemic. This can be achieved through proper projections that can help the governments to take decisions accordingly and create more infrastructure, if required. This projection is particularly important for a country like India which ranks second in population size with high population density in several states. In this paper, a Modified SEIRD (Susceptible-Exposed-Infected-Recovered-Deceased) model was proposed to perform COVID-19 projections for India and its five states having the highest number of total cases. This model also considers asymptomatic infectious patients for the projections. Further, Long Short-Term Memory (LSTM) model was also used in this paper to perform short-term projections. To predict the trend of the pandemic, data up to 15$^{th}$ August 2020 was utilised for experimentation purposes. The results obtained by the proposed Modified SEIRD model and LSTM model were compared for next 30 days. Lower and upper estimates of 90% confidence intervals for the predictions were computed using Student t-test. The effect of different lockdowns imposed in India, was also analysed and modelled using the proposed model.

# Acknowledgement

The authors would like to acknowledge Prof. Daman Saluja and Ms. Apoorva Uboveja (B.R. Ambedkar Centre for Biomedical Research); Dr. Anita Mangla and Dr. Neeru Dhamija (Daulat Ram College), and Dr. Uma Chaudhary (Bhakaracharya College of Applied Sciences) of University of Delhi for the valuable discussion at initial level that created our interest in carrying out this research.